\documentclass[usenatbib]{mn2e}

\usepackage{graphicx}

\title[Dust-eliminated near-infrared spectra of quasars]
{The dust-eliminated shape of quasar spectra in the near-infrared: a
hidden part of the big blue bump}

\author[Kishimoto, Antonucci \& Blaes]
{Makoto Kishimoto$^{1}$\thanks{E-mail: mk@roe.ac.uk}, Robert
Antonucci$^{2}$ and Omer Blaes$^{2}$\\ $^{1}$Institute for Astronomy,
University of Edinburgh, Blackford Hill, Edinburgh EH9 3HJ, UK\\
$^{2}$Physics Department, University of California, Santa Barbara, CA
93106, USA}

\begin{document}

\date{accepted by MNRAS on 5 September 2005}

\pagerange{\pageref{firstpage}--\pageref{lastpage}} \pubyear{2005}

\maketitle

\label{firstpage}

\begin{abstract}

The near-infrared shape of the big blue bump component in quasar
spectra has been essentially unknown. It usually cannot be observed
directly, due to the strong hot dust emission which dominates quasar
spectra longward of $\sim$1$\mu$m.  However this is quite an important
part of the spectrum theoretically. At least bare disk models provide
quite a robust prediction for the overall continuum shape in the
near-infrared. Self-gravity should become important in the outer,
near-infrared emitting regions of the putative disk, possibly leaving
a signature of disk truncation in the near-infrared.  We propose here
that this important part of the spectrum can be revealed for the first
time by observing polarized flux from normal quasars.  At least in
some polarized quasars, the emission lines are all unpolarized and so
the polarized flux should originate interior to the broad line region,
and therefore also interior to the dust emitting region. This can then
be used to eliminate the dust emission.  We present the results of
near-infrared polarimetry for such three quasars (Ton202, 4C37.43, B2
1208+32).  The data for Ton202 have the highest S/N, and the
near-infrared polarized flux in this case is measured to have quite a
blue shape, $F_{\nu} \propto \nu^{+0.42\pm0.29}$, intriguingly
consistent with the simple multi-temperature black body, bare disk
prediction of $\nu^{+1/3}$.  All these data, although still with quite
low S/N for the other two objects, demonstrate the unique potential of
the technique with future better data.  We also present similar data
for other quasars and radio galaxies, and briefly discuss the nature
of the polarization.

\end{abstract}


\section{Introduction}\label{sec-intro}

The infrared portion of quasar spectra is dominated by the thermal
emission from dust grains.  This domination starts in the
near-infrared (near-IR), at $\sim$1$\mu$m, which is essentially set by the
dust sublimation temperature.  This hot dust emission in the near-IR 
actually hides quite an important part of quasar spectra,
i.e. the long wavelength side of the UV/optical continuum component,
often called the big blue bump (BBB).

The radiative energy output of quasars is dominated by this UV/optical
component.  It is generally thought to be from an accretion disk
around a supermassive black hole (e.g. \citealt{Sh78},
\citealt{Ma83}), but the nature of this putative disk has not been
well understood in many respects
(e.g. \citealt{An88}; \citealt{An99}; \citealt{KB99}).  However, disk
models make a few key predictions for the long wavelength spectra.

It is well known that in a very simple multi-temperature black body
disk, extending infinitely in radius, the spectral shape converges to
$F_{\nu} \propto \nu^{+1/3}$ at long wavelengths.  Even with
sophisticated, relativistic disk atmosphere models
(e.g. \citealt{Hu00}), this limit is essentially reached longward of
$\sim$1 $\mu$m for almost all reasonable black hole masses and
Eddington ratios.  While reprocessing by a flared or warped disk can
result in a redder spectrum, we have quite a robust prediction for the
overall continuum shape of at least a bare disk in the near-IR.
Furthermore, the outer parts of standard disks are known to be
unstable against self-gravity, and this might set an upper limit to
the radial extent of the disk (e.g. \citealt{Go03}).  Since the outer
radii generally contribute to longer wavelength emission, this limit
would produce a change in slope in the spectrum, with a rapid flux
down-turn toward long wavelengths.  Standard Shakura-Sunyaev type disk
models predict that, in the absence of any other additional processes,
the truncation radius would be around a thousand Schwarzschild radii
for conceivable parameters of quasars \citep{Go03}.  This in turn
means that its effect on the emission spectrum would show up toward
and at the near-IR.  Thus the near-IR spectral shape of the BBB
emission is quite important for testing fundamental aspects of disk
models.  Near-IR detection of the outer edge of the disk would also
shed much needed light on the mass and angular momentum supply that
fuels the quasar.

The near-IR part of the BBB has remained unseen directly, because of
the hot dust emission described above --- this is thought to be from
the inner part of a torus-like structure surrounding the nucleus and
the broad emission line region (BLR).  We still do not have the
spatial resolution to isolate the nucleus from this surrounding
putative torus.  Spectral decomposition does not constrain the near-IR
BBB shape well \citep{Ma89}.  To uncover this important part of the
spectrum and investigate it directly, we need to remove the hot dust
emission.  The red light from the host galaxy probably contaminates
the observed spectrum as well.  We argue here that we can remove all
these contaminations by using polarization.

\begin{figure}
 \includegraphics[width=80mm]{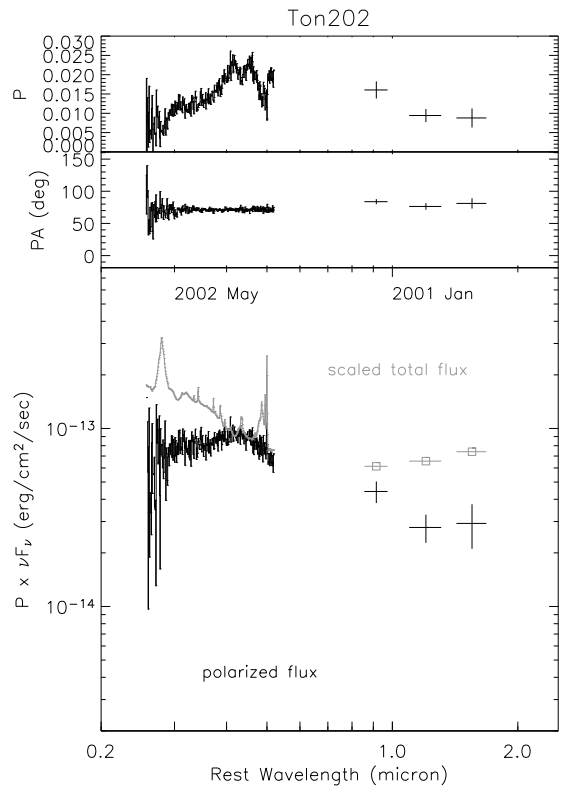} 

 \caption{The near-infrared broad-band polarization measurements of the
 Ton202 taken in Jan 2001, shown with optical spectropolarimetry data
 taken in May 2002 from \citet{Ki03}. The optical and near-infrared
 total flux is scaled to roughly match the red side of the optical
 polarized flux for a clearer comparison.}

 \label{fig-Ton202}
\end{figure}

\begin{figure}
 \includegraphics[width=80mm]{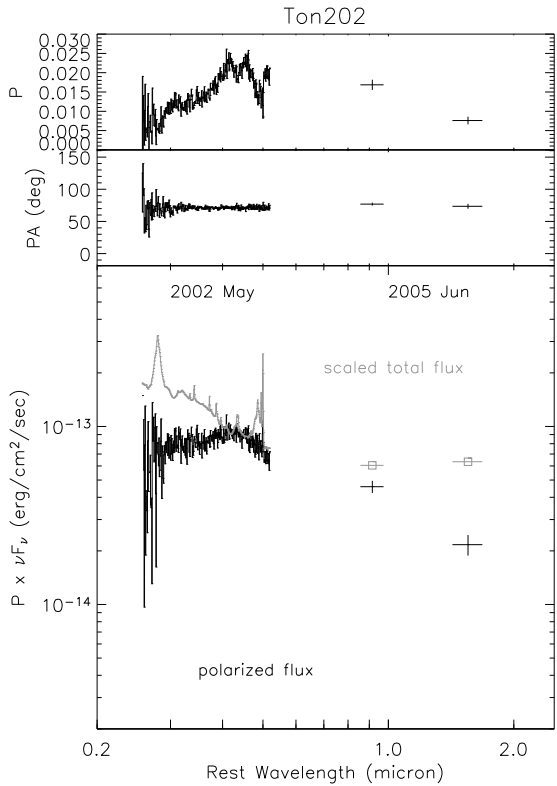} 
 \caption{The same as Fig.\ref{fig-Ton202} but with the 
 near-infrared data taken in 2005.}
 \label{fig-Ton202-2005}
\end{figure}

\begin{figure}
 \includegraphics[width=80mm]{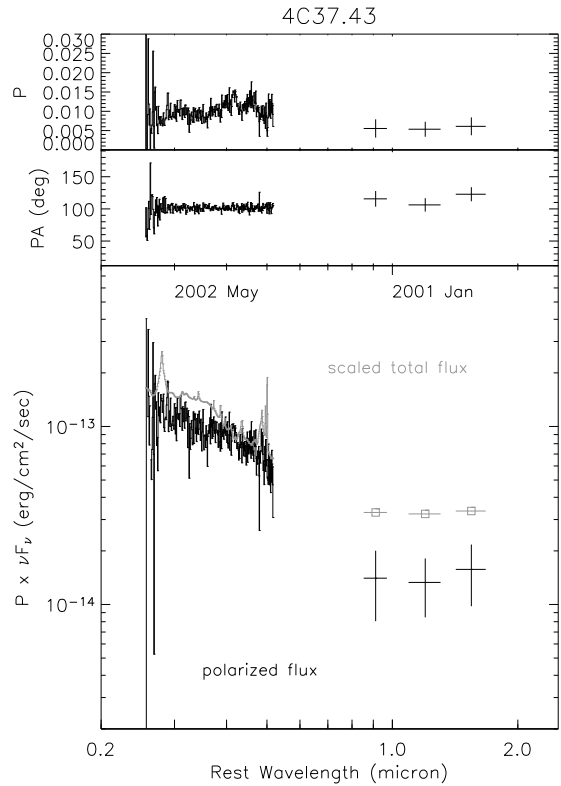} 
 \caption{The same as Fig.\ref{fig-Ton202} but for 4C37.43.}
 \label{fig-4C37.43}
\end{figure}

\begin{figure}
 \includegraphics[width=80mm]{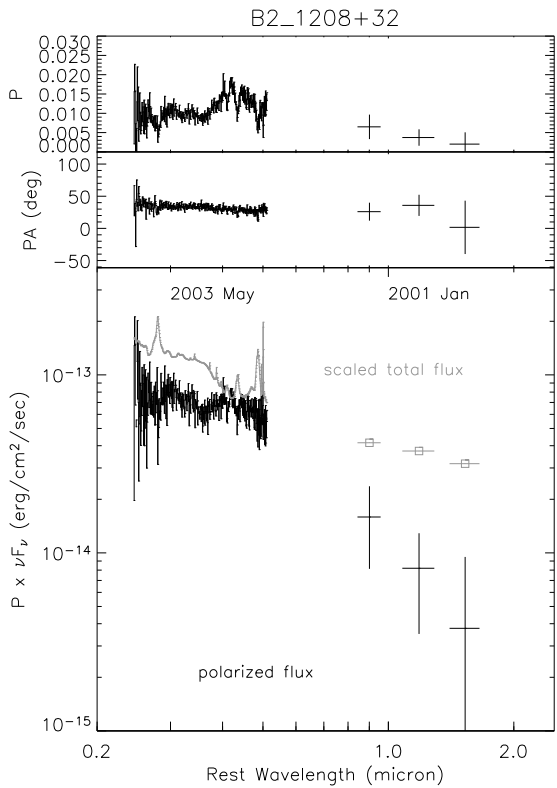} 

 \caption{The same as Fig.\ref{fig-Ton202} but for B2 1208+32. The
 optical spectropolarimetry data were taken in May 2003.}

 \label{fig-B2_1208+32}
\end{figure}

Many normal quasars show a small ($P\sim1$\%) optical polarization
\citep*{St84,Be90}, and there is a strong statistical tendency for the
polarization position angle (PA; E-vector direction) to lie parallel
to the radio structural axis in these quasars (e.g. \citealt{St79,
MS84, RS85}). At least in some (and possibly many) cases, this
polarization seems to be confined to the continuum --- there are
essentially no emission lines in the polarized flux ($P \times F$; see
Figure 31 in \citealt{Ki04}; the optical data for three quasars are
reproduced in Figure~\ref{fig-Ton202}-\ref{fig-B2_1208+32} in this
paper). In this case, the polarized flux is considered to originate
{\it interior to} the BLR. We can then scrape off all the
contaminations coming from the BLR and outer regions just by looking
at the polarized flux.  We have applied this idea in the
near-UV/optical to look for the Balmer edge feature, intrinsic to the
BBB emission but buried under the contamination from the BLR. Indeed
we found the edge feature in absorption --- this is most simply
interpreted as an indication of the optically-thick thermal nature of
the BBB emission (\citealt*{Ki03}; \citealt{Ki04}).

In principle, we should be able to apply the same idea to remove the
dust emission, which arises exterior to the BLR.  We present here the
near-IR polarized flux measurements for three quasars which have no or
very little emission lines in the optical polarized flux.  The data
are still of low S/N or in need for further measurements, but suggest
that our idea seems to work.  We also present near-IR polarization
measurements for other quasars and radio galaxies and briefly discuss
the nature of the polarization.

\begin{table}
  \caption{Log of UKIRT observations in 2001.}
    \begin{tabular}{llccccc}
  \hline
  name /      & other     & $z$   & exp.          & date        &   filter \\
  position    & name      &       & (min)         &             & \\
  \hline

  0837-120    & 3C206     & 0.198 &  8 $\times$ 1 & 15 Jan & $J$ \\
              &           &       & 12 $\times$ 1 & 15 Jan & $J$ \\
              &           &       &  4 $\times$ 3 & 15 Jan & $H$ \\
              &           &       &  8 $\times$ 1 & 15 Jan & $K'$ \\

  1114+445    & PG        & 0.144 & 12 $\times$ 1 & 15 Jan & $J$ \\
              &           &       & 12 $\times$ 1 & 16 Jan & $J$ \\
              &           &       &  4 $\times$ 1 & 15 Jan & $H$ \\
              &           &       &  4 $\times$ 1 & 16 Jan & $H$ \\
              &           &       &  4 $\times$ 1 & 15 Jan & $K'$ \\
              &           &       &  4 $\times$ 1 & 16 Jan & $K'$ \\

  1208+322    & B2        & 0.388 & 12 $\times$ 1 & 16 Jan & $J$ \\
              &           &       &  8 $\times$ 2 & 15 Jan & $H$ \\
              &           &       &  8 $\times$ 1 & 16 Jan & $H$ \\
              &           &       &  4 $\times$ 1 & 15 Jan & $K'$ \\
              &           &       &  8 $\times$ 1 & 15 Jan & $K'$ \\
              &           &       &  8 $\times$ 1 & 16 Jan & $K'$ \\

  1425+267    & Ton202    & 0.366 & 12 $\times$ 2 & 16 Jan & $J$ \\
              &           &       &  8 $\times$ 2 & 16 Jan & $H$ \\
              &           &       &  8 $\times$ 1 & 16 Jan & $K'$ \\
        
  1512+370    & 4C37.43   & 0.371 & 12 $\times$ 2 & 16 Jan & $J$ \\
              &           &       &  8 $\times$ 2 & 16 Jan & $H$ \\
              &           &       &  8 $\times$ 2 & 16 Jan & $K'$ \\
  \hline
  \end{tabular}

  \label{tab-log-2001}
\end{table}

\begin{table}
  \caption{Log of UKIRT observations in 2000.}
  \begin{tabular}{llccccc}
\hline
name/        & other     & $z$   & exp.          & date        &   filter \\
position     & name      &       & (min)         &             & \\ 
\hline

0453+227    & 3C132     & 0.214 & 10$\times$2 & 25 Jan & $K'$ \\ %
0459+252    & 3C133     & 0.278 & 10$\times$1 & 26 Jan & $K'$ \\

0752+258    & OI287     & 0.446 & 10$\times$1 & 24 Jan & $K$ \\ 
            &           &       & 10$\times$1 & 25 Jan & $K'$ \\

0802+243    & 3C192     & 0.060 & 10$\times$2 & 26 Jan & $K'$ \\
0824+294    & 3C200     & 0.458 & 10$\times$3 & 26 Jan & $K'$ \\

0938+399    & 3C223.1   & 0.107 & 10$\times$1 & 24 Jan & $K'$ \\
            &           &       & 10$\times$2 & 25 Jan & $K'$ \\

0945+076    & 3C227     & 0.086 & 10$\times$1 & 24 Jan & $K$ \\ 
            &           &       & 10$\times$2 & 24 Jan & $K'$ \\
            &           &       & 10$\times$1 & 25 Jan & $K'$ \\

0958+290    & 3C234     & 0.185 & 10$\times$2 & 24 Jan & $K'$ \\

1003+351    & 3C236     & 0.101 & 10$\times$2 & 25 Jan & $K'$ \\

1142+318    & 3C265     & 0.811 & 10$\times$4 & 26 Jan & $K'$ \\


1420+198    & 3C300     & 0.270 & 10$\times$4 & 26 Jan & $K'$ \\

1512+370    & 4C37.43   & 0.371 & 10$\times$1 & 25 Jan & $K'$ \\


\hline
\end{tabular}

  \label{tab-log-2000}
\end{table}

\begin{table}
  \caption{Log of UKIRT observations in 2005.}
    \begin{tabular}{llccccc}
  \hline
  name /      & other     & $z$   & exp.          & date        &   filter \\
  position    & name      &       & (min)         &             & \\
  \hline

  1425+267    & Ton202    & 0.366 & 12 $\times$ 6 & 27 Jun & $J$ \\
              &           &       &  8 $\times$ 8 & 27 Jun & $K'$ \\
        
  \hline
  \end{tabular}

  \label{tab-log-2005}
\end{table}

\begin{table*}
\begin{minipage}{110mm}
  \caption{Observation of unpolarized stars with de-focusing.}
  \begin{tabular}{lccccccccccccccccccccccc}
\hline
name & UT date & band & mag & P(per cent) & PA(\degr) & aperture$^a$\\

\hline
SAO82926$^b$ & 2005 Jul 8& $J$ &  9.33 & $0.09\pm0.06$ & $157\pm 21$ & 2.7\\
&  & $K'$ &  9.12 & $0.11\pm0.04$ & $ 29\pm 12$ & \\

SAO82926$^c$ & 2005 Jul 8& $J$ &  9.33 & $0.24\pm0.05$ & $ 99\pm  6$ & 3.6\\
&  & $K'$ &  9.11 & $0.24\pm0.05$ & $ 59\pm  6$ & \\

CMC604707$^c$ & 2001 Jan 16& $J$ &  9.25 & $0.25\pm0.05$ & $168\pm  5$ & 3.6\\
&  & $H$ &  8.87 & $0.09\pm0.03$ & $105\pm  9$ & \\
&  & $K'$ &  8.81 & $0.12\pm0.04$ & $124\pm  10$ & \\

CMC604919$^d$ & 2001 Jan 15& $J$ &  7.99 & $0.14\pm0.02$ & $ 92\pm  5$ & 7.5\\
&  & $H$ &  7.37 & $0.19\pm0.01$ & $ 100\pm  2$ & \\
&  & $K'$ &  7.23 & $0.14\pm0.02$ & $ 82\pm  4$ & \\

\hline
\end{tabular}
\\Note. $^a$The synthetic aperture diameter in arcsec 
for polarization measurements. The apertures for the photometric 
magnitudes quoted are larger than these. $^b$Observed with a small de-focusing.
$^c$Observed with a further de-focusing. $^d$Observed with a large
de-focusing.

  \label{tab-unpol}
\end{minipage}
\end{table*}

\begin{table*}
\begin{minipage}{150mm}

  \caption{Results of polarization measurements. The polarization
degree has not been debiased. The magnitude-flux conversion is from
\citet{TV05}. The flux $F_{\lambda}$ is in units of
erg/cm$^2$/sec/\AA. The flux calibration uncertainty is estimated to
be typically $\sim$5\%, and this is much larger than the statistical
error in the flux measurements for all the objects.}

  \begin{tabular}{lccccccccccccccccccccccc}
\hline
name & z & $E(B-V)$ & year & band & mag & $F_{\nu}$ (mJy) & $F_{\lambda}$ & P(per cent) & PA(\degr)\\

\hline
\multicolumn{8}{l}{quasars with unpolarized broad lines}\\
Ton202 & 0.366 & 0.019 & 2001 & $J$ & 15.35 &  1.13 & 2.18E-16 & $1.60\pm0.22$ & $ 83.8\pm  4.0$\\
 &  &  &  & $H$ & 14.53 &  1.60 & 1.82E-16 & $0.94\pm0.17$ & $ 76.4\pm  5.2$\\
 &  &  &  & $K'$ & 13.67 &  2.34 & 1.56E-16 & $0.88\pm0.25$ & $ 81.1\pm  8.1$\\

 &  &  & 2005 & $J$ & 15.36 &  1.12 & 2.16E-16 & $1.69\pm0.13$ & $ 76.9\pm  2.2$\\
 &  &  &  & $K'$ & 13.84 &  2.00 & 1.33E-16 & $0.76\pm0.10$ & $ 73.5\pm  3.8$\\

B2 1208+32 & 0.388 & 0.017 & 2001 & $J$ & 15.47 &  1.01 & 1.95E-16 & $0.65\pm0.32$ & $ 26.0\pm 14.0$\\
 &  &  &  & $H$ & 14.85 &  1.20 & 1.36E-16 & $0.37\pm0.21$ & $ 35.8\pm 16.4$\\
 &  &  &  & $K'$ & 14.29 &  1.32 & 8.76E-17 & $0.20\pm0.31$ & $  1.6\pm 41.5$\\

4C37.43 & 0.371 & 0.022 & 2001 & $J$ & 15.44 &  1.04 & 2.00E-16 & $0.56\pm0.24$ & $115.6\pm 12.2$\\
 &  &  &  & $H$ & 14.72 &  1.35 & 1.53E-16 & $0.54\pm0.19$ & $106.2\pm 10.5$\\
 &  &  &  & $K'$ & 13.95 &  1.81 & 1.21E-16 & $0.61\pm0.23$ & $122.8\pm 11.0$\\

 &  &  & 2000 & $K'$ & 13.89 &  1.90 & 1.27E-16 & $0.77\pm0.30$ & $ 96.4\pm 11.9$\\

\\ \multicolumn{8}{l}{other quasars}\\
3C227 & 0.086 & 0.026 & 2000 & $K'$ & 12.44 &  7.22 & 4.81E-16 & $0.42\pm0.06$ & $ 22.0\pm  4.0$\\
 &  &  &  & $K$ & 12.31 &  7.71 & 4.78E-16 & $0.41\pm0.11$ & $ 28.5\pm  7.8$\\

OI287 & 0.444 & 0.061 & 2000 & $K'$ & 14.11 &  1.56 & 1.04E-16 & $2.26\pm0.40$ & $126.9\pm  5.4$\\
 &  &  &  & $K$ & 13.87 &  1.82 & 1.13E-16 & $2.05\pm0.44$ & $128.8\pm  5.8$\\

Q1114+445 & 0.144 & 0.016 & 2001 & $J$ & 14.38 &  2.76 & 5.33E-16 & $1.17\pm0.13$ & $ 83.6\pm  3.1$\\
 &  &  &  & $H$ & 13.57 &  3.87 & 4.39E-16 & $0.70\pm0.16$ & $ 97.9\pm  6.4$\\
 &  &  &  & $K'$ & 12.47 &  7.00 & 4.67E-16 & $0.96\pm0.12$ & $ 88.0\pm  3.5$\\

3C206 & 0.198 & 0.045 & 2001 & $J$ & 14.49 &  2.49 & 4.80E-16 & $0.25\pm0.19$ & $107.9\pm 21.7$\\
 &  &  &  & $H$ & 13.85 &  2.99 & 3.39E-16 & $0.42\pm0.19$ & $ 90.8\pm 12.8$\\
 &  &  &  & $K'$ & 12.83 &  5.08 & 3.39E-16 & $0.21\pm0.16$ & $130.4\pm 21.8$\\

\\ \multicolumn{8}{l}{narrow-line radio galaxies}\\
3C234 & 0.185 & 0.019 & 2000 & $K'$ & 12.88 &  4.84 & 3.22E-16 & $4.60\pm0.11$ & $159.2\pm  0.7$\\

3C265 & 0.811 & 0.023 & 2000 & $K'$ & 16.12 &  0.24 & 1.61E-17 & $4.71\pm1.49$ & $ 10.1\pm  9.0$\\

3C223.1 & 0.107 & 0.017 & 2000 & $K'$ & 12.91 &  4.70 & 3.13E-16 & $5.24\pm0.15$ & $108.4\pm  0.8$\\

3C132 & 0.214 & 0.482 & 2000 & $K'$ & 14.33 &  1.27 & 8.49E-17 & $1.08\pm0.70$ & $ 48.8\pm 18.2$\\

3C133 & 0.278 & 0.949 & 2000 & $K'$ & 14.97 &  0.70 & 4.68E-17 & $6.96\pm1.38$ & $ 93.6\pm  6.3$\\

3C192 & 0.060 & 0.054 & 2000 & $K'$ & 12.86 &  4.94 & 3.29E-16 & $0.03\pm0.17$ & $102.8\pm 90.0$\\

3C200 & 0.458 & 0.040 & 2000 & $K'$ & 15.33 &  0.50 & 3.34E-17 & $2.25\pm1.05$ & $127.4\pm 13.0$\\

3C236 & 0.101 & 0.011 & 2000 & $K'$ & 12.89 &  4.81 & 3.20E-16 & $0.48\pm0.19$ & $ 54.7\pm 11.4$\\

3C300 & 0.270 & 0.035 & 2000 & $K'$ & 15.32 &  0.51 & 3.40E-17 & $1.27\pm0.85$ & $ 61.0\pm 19.3$\\

\hline
\end{tabular}

  \label{tab-pol}
\end{minipage}
\end{table*}

\section{Observations}\label{sec-obs}

We observed several quasars listed in Table~\ref{tab-log-2001} in the
second half nights of 15 and 16 Jan 2001 (UT) with UKIRT.  The sample
was selected mainly based on the known polarization properties: they
are all known to be polarized in the optical with 1-2\% level except
for 3C206 \citep{St84,Be90}, and all have previous polarization
measurements at $K$ band except for B2 1208+32 although with rather
large uncertainties \citep{SZ91}. They are lobe-dominant radio sources
except for Q1114+445 which is radio quiet.  We used the camera UFTI
with the polarimetry module IRPOL2. The platescale was $0.''0909$ per
pixel.  Seeing was estimated to be $\sim 1.''2$ FWHM (or slightly
worse) in most of the first night, while it was $\sim 0.''8$ FWHM in
the rest of the run, based on the image size of these quasars.

The data were reduced firstly with the software ORACDR where bad
pixels were masked, dark frame subtracted, and the frames were divided
by a flat-field which was constructed from sky frames at four
waveplate positions (we averaged these four flat-fields from four
waveplate positions after checking consistency), and frames with three
dithering positions were registered and combined. Then synthetic
aperture photometry was implemented using IDL in each o-ray and e-ray
image at four waveplate positions to produce Stokes $I$, $Q$, and $U$
parameters.  This was done with various aperture radii and optimal
aperture size was searched to minimize the polarization measurement
uncertainty.  We adopted a $1.''8$ diameter aperture for the data with
the worse seeing size, and $1.''3$ diameter for the rest of the data.
For flux measurements, a larger aperture was chosen: for the data with
the worse seeing we adopted $5.''5$ diameter, while for the rest
$3.''6$ diameter. The sky level was determined at the annulus with
inner and outer diameters being $5.''5$ and $7.''3$ for the worse
seeing data and $3.''6$ and $5.''5$ for the rest (both for
polarization and flux measurements).

We generally repeated a set of polarization measurements a few times
with the same filter at four waveplate positions. The number of
repeated sets is shown in the exposure time column of
Tables~\ref{tab-log-2001}-\ref{tab-log-2005}. The Stokes parameters
measured in each set were combined by taking the weighted average.
Polarization position angles were calibrated using the polarization
standard star HDE 283809 with an uncertainty of 3\degr.

We also observed several quasars and radio galaxies listed in
Table~\ref{tab-log-2000} in the second half nights of 24, 25, and 26
Jan 2000 (UT) with the same instruments on UKIRT.  Seeing is estimated
to be $\sim 1''.4$ for the first night and $\sim 0.''9$ for the second
and third nights.  During this run in 2000, we occasionally had a
guiding problem due to the vignetting of the field of view by the
waveplate holder.  This made the images of some objects slightly
elongated or distorted.  For some objects with this guiding problem,
we used our own IDL scripts to find the best registration of the
images.  Otherwise the data were reduced in the same manner as above.
We used rather large synthetic apertures for all the frames in this
run, namely the same apertures as those for the worse seeing data in
2001 described above ($1.''8$ and $5.''5$ for polarization and flux
measurement, respectively, with the background determined at the
annulus between $5.''5$ and $7.''3$, all in diameter).  For 3C192,
3C236, 3C223.1 which were quite extended, we used larger apertures for
flux measurements ($7.''3$ for 3C192 and $9.''1$ for 3C236 and
3C223.1) and background determinations ($7.''3$/$9.''1$ and
$9.''1$/$10.''9$ annuli).

The data taken in 2001 were flux-calibrated using the standard star
FS21, and a consistent calibration was obtained for the data in 2000
using the standard star FS134.  The accuracy of the flux calibration
is estimated to be typically $\sim$5\%, based on the observation of
another flux standard star and also on the comparison of the frames
for the objects with multiple data sets and/or observing dates. 

Finally we also obtained data for one quasar, Ton 202, through the
UKIRT service programme on 27 Jun 2005, kindly helped by Chris
Davis. The observations are listed in Table~\ref{tab-log-2005}.
Seeing is estimated to be $\sim0.''6$. We have used a $0.''9$ aperture
to measure polarization, $3.''6$ for flux measurements, with
background determined between $3.''6$ and $5.''5$ (all in
diameter). The flux standard star was GSPC P272-D. A similar
estimation as above suggests the flux uncertainty to be $\sim$5\% or
better.

Instrumental polarization was checked in the service program in 2005
by observing the star SAO82926 which is known to be unpolarized in the
optical ($0.06\pm0.04$\%; \citealt{KR86}).  The data were obtained on
2005 Jul 8, again kindly helped by Chris Davis. In 2001 we also
observed the stars with high proper motions, CMC604919 and CMC604707,
for which we expect very low intrinsic polarization. These
observations are summarized in Table~\ref{tab-unpol}.  For these
bright stars, we needed to de-focus the telescope to avoid having
counts in a non-linear regime.  Accordingly large synthetic apertures
were used for polarization measurements.  The results seem to show
that the instrumental polarization is smaller than $\sim$0.2\% even
for those de-focused data requiring large apertures, and thus it may
be as good as $\la 0.1$\%.  We hope to obtain fainter unpolarized
stars without any defocusing, in order to better characterize the
instrumental polarization.

\begin{table}

  \caption{Spectral index $\alpha$ of polarized flux ($F_{\nu} \propto
  \nu^{\alpha}$).  The optical slope is measured at the rest
  wavelength interval of 4000\AA\ - 4900\AA\ (4700\AA\ for 4C37.43) as
  described in \citet{Ki03}, by fitting a broken power law with a
  break at 4000\AA. The near-IR slope is measured using the data at $J/H/K'$
  ($J/K'$ for Ton202 data taken in 2005). The optical-near-IR slope is
  measured by first measuring the average optical polarized flux at
  the interval quoted above and then fitting a power law for this and
  near-IR data at $J/H/K'$ (or $J/K'$) allowing for the uncertainty in the
  absolute flux scale in the optical.}

  \begin{tabular}{lccccccccccccccccccccccc}
\hline
name & optical & optical-nIR & nIR\\

\hline
Ton202     & $-0.54\pm0.08$ & $+0.05\pm0.16$ & $+0.15\pm0.60$  \\

Ton202$^a$ &  & $+0.08\pm0.12$ & $+0.42\pm0.29$ \\

B2 1208+32 & $-0.54\pm0.11$ & $+1.15\pm0.60$ & $+1.6 \pm2.2 $ \\

4C37.43    & $+0.06\pm0.23$ & $+0.88\pm0.29$ & $-1.2 \pm1.1 $ \\




\hline
\end{tabular}
\\Note. $^a$The measurements using the near-IR data taken in 2005.

  \label{tab-slope}
\end{table}

\section{Results and discussions}  

\subsection{Quasars with unpolarized broad lines}

\subsubsection{Results}

The primary objects here are the ones which have been found to have
unpolarized broad lines \citep{Ki03,Ki04}, namely Ton202, 4C37.43, and
B2 1208+32 in our present sample. The results of the near-IR
polarimetry for these objects, as well as for others, are summarized
in Table~\ref{tab-pol}.  Note that the polarization degree $P$ has not
been debiased and so the observed Stokes parameters can be restored
directly from $P$ and position angle (PA), and that the biases for
those with good polarization detections are always small \citep{SS85}.
The statistical errors in the flux measurements are much smaller than
the flux calibration error for all the objects and therefore are not
shown in Table~\ref{tab-pol} (the fractional statistical error in
total flux $\sigma_F / F$ is roughly of the order of $\sigma_P$).

Figure~\ref{fig-Ton202} and \ref{fig-Ton202-2005} show the spectral
energy distribution of Ton202 from the optical to near-IR in polarized
flux as well as in total flux, the former with the near-IR data taken
in 2001 and the latter with those taken in 2005.
Figure~\ref{fig-4C37.43} and \ref{fig-B2_1208+32} present the same
spectra for 4C37.43 and B2 1208+32, respectively.  The optical
spectropolarimetry data are from \citet{Ki03} and \citet{Ki04}.  (Note
that the optical polarized flux shown is actually a rotated Stokes
flux, i.e. the unnormalized Stokes parameter $Q$ with the reference
axis at each object's polarization PA through the optical band. This
quantity does not suffer from the positive bias of polarization
degrees, and represents the polarized flux accurately when a single PA
for the object can be determined with a high S/N.) All the flux in the
Figures has been corrected for the Galactic reddening using the values
of $E(B-V)$ in Table~\ref{tab-pol} and the reddening curve of
\citet{Ca89} with $R_V = 3.1$.

The flux measurements in the optical spectroscopic data have some
uncertainty from slit loss and seeing change. Therefore, for Ton202
and 4C37.43, we have scaled the flux by comparing the flux of the
nuclear [OIII]$\lambda$4959 and $\lambda$5007 lines in our spectra
with those of \citet{Sh03} (the [OIII] flux is known to be spatially
extended in these objects; \citealt{SM87}). They used the continuum
level from wide-slit spectra to scale their narrow-slit spectra, which
are considered to contain essentially only the nuclear [OIII] flux for
these two objects. (For our 4C37.43 data, we re-extracted the spectrum
with a smaller window to measure the nuclear [OIII] flux.)  The
resulting nuclear [OIII] flux for each object was further checked with
the nuclear [OIII] flux measurements of \citet{SM87}.  We adopted a
scaling factor of 1.33 for the Ton202 data, and 1.41 for 4C37.43, and
estimate the flux calibration uncertainty of $\sim$ 10\% based on the
calibration uncertainty of Shang et al. data and slight discrepancies
between the [OIII] flux in our scaled spectra and that of Stockton \&
MacKenty. However, the uncertainty in this scaling might be somewhat
larger than this, especially for 4C37.43 which has quite a large
extended [OIII] flux. For B2 1208+32, we did not make any adjustment
due to the unavailability of such data.

For one of these three objects, Ton202, we detected near-IR
polarization with S/N larger than 3 in three bands in 2001, and the
data obtained in 2005 in $J$ and $K'$ bands show consistent polarization
with much higher S/N (we note that $K'$-band total flux might have
varied - see Table~\ref{tab-pol}).  For the other two objects,
especially for B2 1208+32, the measurements are still of low S/N, but
the data show coherent PAs in multiple bands.  The PA is roughly the
same as those measured in the optical in all three objects, and are
all roughly parallel (within about 30\degr) to the radio structural
axis PAs which are 53\degr, 109\degr, and 3\degr\ for Ton202, 4C37.43,
and B2 1208+32, respectively (see \citealt{Ki04} for references).  The
polarization degree in the near-IR is generally lower than in the
optical, and it seems to be decreasing toward longer wavelengths in
the near-IR at least in the highest S/N case of Ton202.

\subsubsection{Interpretations}

As we claimed in the previous papers \citep{Ki03,Ki04}, the optical
polarization in these quasars is likely due to electron scattering
{\it interior to} the BLR, essentially because there are no emission
lines seen in the optical polarized flux.  The polarization mechanism
in the near-IR does not seem to be different from the optical at least
based on the same PA observed.  If this is the case, the simplest
interpretation is that the polarized flux spectrum over the
optical-near-IR region is an electron-scattered copy of the emission
interior to the BLR, revealing the spectral shape of the big blue bump
emission. Note that the scatterers would be electrons and not dust
grains, since the scattering region would be within the dust
sublimation radius. If this simplest interpretation is correct, then
the polarized flux gives the actual measurement of the near-IR
shape of the big blue bump for the first time.  This idea is
consistent with the tendency of the polarization degree decreasing
toward longer wavelengths while the total flux is turning up due to
the onset of hot dust emission longward of 1$\mu$m.

We have measured the slope of the polarized flux in different
wavelength ranges as shown in Table~\ref{tab-slope}.  The measurements
show that the polarized flux over the optical to near-IR looks quite
blue compared to the polarized flux shape in the optical alone.  The
improved data for Ton202 taken in 2005 show that the polarized flux
slope in the near-IR alone, $F_{\nu} \propto \nu^{+0.42 \pm 0.29}$, is
also quite blue compared to the shape in the optical.  Intriguingly,
the blue slope is consistent with the long wavelength limit of simple
multi-temperature black-body disks, $F_{\nu} \propto \nu^{+1/3}$.
However, the measurements for the other two objects (4C37.43 and B2
1208+32) are still of low S/N, and any general definitive statements
have to wait for future data, including $z$-band and red-side optical
data for Ton202.  Nevertheless, these results, even the present Ton202
data alone, at least suggest that the near-IR polarized flux
measurements have the potential to remove the hot dust emission and
reveal the underlying intrinsic spectral shape there.

\subsubsection{Comparison with disk models}

\begin{figure}
 \includegraphics[width=80mm]{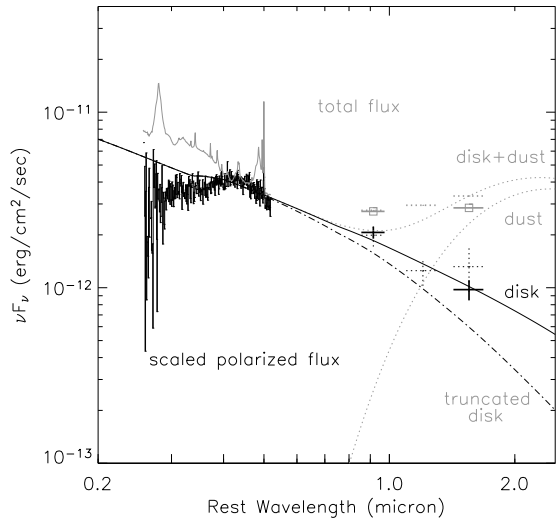} 

 \caption{Comparison of the observed total and polarized flux for
 Ton202 with a disk model based on the atmosphere calculations of
 \citet{Hu00}.  The observed data are the same as shown in
 Figure~\ref{fig-Ton202} and Figure~\ref{fig-Ton202-2005}, and the
 near-IR data taken in 2001 are shown in dotted lines. All the
 polarized flux has been scaled up by a factor of 45 to match the
 total flux at the red side in the optical. The solid smooth curve is
 the disk model, while the dash-dot curve represents the same model
 truncated at the nominal self-gravitating radius. The lower dotted
 curve represents hot dust emission, and the upper dotted curve is the
 sum of this dust emission and the disk model without the
 truncation. See text for more details.}

 \label{fig-diskdust}
\end{figure}

To illustrate the potential of this method for testing emission models
for the optical/near-IR portion of the spectrum, we have computed
accretion disk spectra and a simplified dust model appropriate for
Ton202.  For the disk models, we adopted a fixed black hole mass of
$M_{\rm BH}=8.4\times10^8 M_{\sun}$ and an Eddington ratio of
$L/L_{\rm Edd}=0.15$, values derived from the H$\beta$ line width and
optical continuum luminosity using the \citet{Ka05} relation.  We then
computed 60 spectra for relativistic accretion disks with the non-LTE
atmospheres of \citet{Hu00}.  The parameters of these spectra ranged
over five different black hole spins ($a/M=0$, 0.5, 0.9, 0.99, and
0.998), six different disk inclinations ($\cos i=0.01$, 0.2, 0.4, 0.6,
0.8, and 1), and two different values of the Shakura-Sunyaev alpha
viscosity (0.01 and 0.1).  After scaling the optical polarized flux up
by a factor 45 in order to match the long wavelength optical total
flux in the continuum, we then fit these models to the {\it optical
polarized flux} at $\lambda\lambda 3000-4500$\AA in the rest frame,
{\it without taking account of the near-IR data}.  The best fit disk
spectrum turned out to have $a/M=0.99$, $\cos i=0.8$, and
$\alpha=0.1$.  These parameters should not be taken too literally,
however, as our model grid was quite coarse and we did not account for
any errors in the assumed black hole mass and Eddington ratio.
Figure~\ref{fig-diskdust} compares the disk spectrum to the total and
polarized flux data (the latter is scaled up by a factor 45).  Note
that the disk spectrum sails right through the near-IR polarized flux
points, even though no attempt was made to incorporate them in the
fit.

It is important to note that this long wavelength behaviour is robust
in these bare disk models: the slope of the model spectrum measured
between 1$\mu$m and 2$\mu$m is $F_{\nu} \propto \nu^{+0.19}$, which is
quite close to the canonical $\nu^{+1/3}$ behaviour, and this does not
sensitively depend on any parameters above, including $M_{\rm BH}$
and $L/L_{\rm Edd}$.  At shorter (optical/ultraviolet) wavelengths,
the models depend much more sensitively on parameters, and this
particular model does not seem to reproduce the observed spectrum well
in this region.

To illustrate the contribution of the hot dust emission in the near-IR
total flux, a black body radiation with temperature $T=1500$K is
included in Figure~\ref{fig-diskdust}. We normalized the black body
spectrum to match the observed $K'$ total flux (from 2001 data; note
the slight difference from 2005 data), when summed with the disk
spectrum.  As we can see, this dust+disk spectrum underpredicts the
total flux at $J$ band (rest $\sim$0.9$\mu$m). The remaining part of
the flux may well be the host galaxy light. In fact, the host
luminosity quoted in \citet{Bo05} for Ton202 is roughly consistent
with the residual flux.  Thus, as we expected, the polarized flux
spectrum seems to successfully remove this component as well.

The disk model actually becomes nominally self-gravitating at a radius
of 330~$GM/c^2$, and Figure~\ref{fig-diskdust} also shows the spectrum
of the model truncated at this radius.  The spectrum becomes much
bluer longward of a break at around 1$\mu$m, corresponding to that
truncation radius.  The spectrum deviates from the near-IR polarized
flux data, although it is probably premature to rule out a truncated
disk given the uncertainties in the actual location of the
self-gravitating radius and the S/N of the data.  This issue should be
explored further with future data for many objects.

\subsubsection{Error sources}

A potential error source in this measurement would be that another
polarization component with a perpendicular PA can reduce the observed
polarized flux without changing the observed PA (note that if the
component is not at a perpendicular PA, it would be recognizable in PA
rotation).  A possible concern in this regard would be the scattered
dust emission in the near-IR, i.e. the hot dust emission from the
putative torus scattered in an outer region, which would be
perpendicularly polarized and might become important in the long
wavelengths where dust emission dominates in the total flux.  However,
we would be able to discard such possibilities if we find that the
slope down-turn toward long wavelengths (i.e. becoming much bluer than
the optical slope) actually starts at much shortward of 1 $\mu$m in
the rest frame, since the effect from dust emission should not be
significant at such short wavelengths.  Therefore it is important to
observe polarized flux shape at shortward of 1 $\mu$m in the rest
frame as well as at longer wavelengths.  We can also minimize the risk
of the confusion from other polarization components by observing as
many objects as possible to see if the same behaviour of the polarized
flux with a constant PA across the optical to near-IR is seen
systematically.

Another potential error source would be the possible existence of
molecular spectral lines in the near-IR BBB spectrum, since this part
of the spectrum would be from a fairly low temperature region. The
spectrum eventually needs to be properly modelled, and near-IR
spectropolarimetry would be valuable to observe actual details
there. However, rough measurements with broad band filters are
certainly of higher priority at this stage.

\subsection{Other objects}

\subsubsection{Broad-line objects}

The broad line radio galaxy 3C227 is a slightly different case from
the quasars described above. This is the object where broad lines are
polarized, and at a different PA from that of its continuum
polarization, and those PAs are not aligned with the radio structural
axis \citep{Co99} which is at 85\degr\ \citep{Mo93}.  Our near-IR
polarization measurement seems to be in line with the optical
measurements, showing roughly consistent PA at lower $P$ ($P$ is
decreasing toward longer wavelengths in the optical).  The polarized
flux slope over the optical to near-IR is roughly $F_{\nu} \propto
\nu^{-0.7}$ measured from the flux at 5500\AA\ and $K'$ band.

The implication of the different continuum polarization PA from the
line PA is that the dominant scattering region is slightly outside the
BLR, but quite close to it. The BLR as seen from the scattering region
is resolved, as opposed to the continuum source.  This is actually
thought to be the case in many Seyfert 1 galaxies \citep{Go94,Sm02}.
In this case the near-IR polarized flux might contain some scattered dust
emission, depending on the location of the scatterers relative to the
inner part of the putative obscuring torus, the source of hot dust
emission.

\subsubsection{Narrow-line objects}

As is previously established, type 2 objects (Seyfert 2 galaxies and
narrow-line radio galaxies or NLRGs) show distinct polarization
features from these type 1 objects (\citealt{An93} and references
therein). The continuum and broad lines are polarized at the same PA,
and it is perpendicular to the radio structural axis.  In these cases,
the scattering region is thought to be located well outside of the
BLR, as confirmed with HST polarization images
(e.g. \citealt{Hu99,Ki01}).  In our sample, 3C234 and 3C265 are in
this category.  When the scattering region is of a scale much larger
than the inner part of the putative torus, the near-IR polarized flux
would almost certainly contain scattered dust emission, so that it
would possibly have an up-turn longward of 1$\mu$m, depending on the
nature of the scatterers (electrons vs dust grains).

Our near-IR polarimetry result for 3C234 confirms that of \citet{SZ91}
with a higher S/N (Table~\ref{tab-pol}), and is also consistent with
the near-IR spectropolarimetric results of \citet{Yo98}.  The PA is
the same as that in the optical, suggesting that the same scattering
mechanism is responsible for the near-IR polarization.  The near-IR
polarized flux of 3C234 in relation to other wavelengths has been
discussed in detail by \citet{Yo98} (see also \citealt{Ki01}, who 
quote the polarized flux slope of $F_{\nu} \propto \nu^{-1.6}$ over
the optical to near-IR).  They suggested that the polarized flux from
dichroism (i.e. selective absorption by aligned dust grains), which is
penetrating through a large amount of column and thus is significantly
reddened, can start to contribute in the near-IR.  This might also
occur in other type 2 objects.  However, note that this additional
dichroic polarization component should have the PA which is either
exactly parallel or perpendicular to that of the scattered
polarization component (though in the latter case the net polarized
flux will decrease) in order for the observed PA to be constant across
the relevant wavelengths.

Our near-IR polarimetry result for 3C265 seems to be in line with the
optical results \citep{JE91,diS96}.  The polarization PA at $K'$ band
is roughly the same as that in the optical but is slightly rotated
(even more perpendicular to the radio axis at PA 106\degr;
\citealt{Fe93}), and the polarization degree at $K'$ band is a little
lower than in the optical.  The slope of the polarized flux derived
from the flux at 5500\AA\ and $K'$ is roughly $F_{\nu} \propto
\nu^{-1.3}$.  Note that the object is at $z=0.811$ so our $K'$
measurement is at the rest wavelength of $\sim1.2\mu$m.

The optical scattered light in NLRGs can further be obscured by
kpc-scale dust lanes. In this case we would see a strong total flux
up-turn in the near-IR toward longer wavelengths with high
polarization.  \citet{AB90} suggested that this is the case in 3C223.1
and also in Cen A.  We confirm the near-IR polarization measurements
of \citet{AB90} for 3C223.1 with a much higher S/N
(Table~\ref{tab-pol}).  We observed several other NLRGs (3C132, 3C133,
3C192, 3C200, 3C236, 3C300; mainly selected by their red $H-K$ and/or
$J-H$ colour) to look for a similar high polarization at $K$ band, but
we did not find such a case (Table~\ref{tab-pol}) except for 3C133.
The observed polarization of 3C133 might not be due to
narrow-line-region-scale scattering, since the observed PA is not
perpendicular but parallel to the radio jet axis which is at PA
107\degr\ \citep{La81}, and a significant part of the polarization may
well be an interstellar polarization (ISP) in our Galaxy (PA is
similar to that of the ISP in the field around the object
[\citealt{He00}], and it is possible for the $K$-band ISP to be
$\sim$2\% for $E(B-V) \sim 1$mag [\citealt{Wh92}]).  We note that
three of these NLRGs (3C132, 3C200, 3C236) are known to have optically
dull spectra and classified as low excitation galaxies or LEG
\citep{La94,JR97}.

\subsubsection{More broad-line objects and type 1 vs type 2}

The difference in the dominant scattering observed in type~1s and
type~2s seems to be that in the former the scattering is in an
equatorial region close to the nucleus (leading to a PA parallel to
the radio axis with broad lines occasionally polarized in a different
PA), while in the latter the scattering is in a polar region far from
the nucleus (leading to a perpendicular PA both for broad lines and
continuum).  The former polarization would dominate when the nucleus
is directly seen, or at least seen with less obscuration.  These two
different scattering regions have been suggested and discussed by
\citet{Sm04,Sm05} for Seyfert 1 galaxies.  A further, important
point is that the size scale of the equatorial scattering region
relative to the BLR appears to be different in different type 1 objects
as discussed in \citet{Ki04}.

In the case of the broad line radio galaxy OI287, the optical
spectropolarimetry shows that the broad lines and continuum are
rather highly polarized with the same $P$ and PA, and the narrow lines
do not appear to be polarized \citep{GM88}.  The PA is parallel to the
radio axis PA which is 146\degr\ \citep{UA88}, and thus the dominant
scattering region seems to be in an equatorial region, but outside of
the BLR and interior to the narrow-line region.  The result of our
near-IR polarimetry is consistent with that of \citet{SZ91}. The
near-IR PA is roughly the same as the optical PA, and the polarization
degree is significantly lower in the near-IR than in the optical.  The
optical to near-IR polarized flux slope is roughly $F_{\nu} \propto
\nu^{-0.8}$ from the flux at 5500\AA\ and $K'$ band.

The case of the radio-quiet QSO 1114+445 is similar to type 2 objects
in that its broad lines and continuum in the optical are polarized in
the same way \citep{Sm93}. The narrow lines are also polarized in the
same way, which is in contrast to the case of OI287, suggesting that
the scattering is exterior to the narrow line region \citep{Sm93}.
Our near-IR polarization measurements at three bands confirm and
extend the result of \citet{SZ91} at $K$ band.  The polarization
degree in the near-IR ($\sim1$\%) is lower than that in the optical
(decreasing from $\sim$3\% to $\sim$2\% toward longer wavelengths).
The PA of the near-IR polarization is essentially the same as that in
the optical, suggesting the same origin.  The polarized flux shape is
roughly $F_{\nu} \propto \nu^{-0.1}$ if measured from the continuum
level at 5500\AA\ and the flux at $K'$ band (or $F_{\nu} \propto
\nu^{+0.3}$ from 5500\AA\ and $J$ band; the near-IR polarized flux
might have an up-turn toward longer wavelengths).

Finally, our near-IR polarization measurements for the quasar 3C206 do
not match the result of \citet{SZ91} which showed $P \sim 2$\% at $K$,
though we note that the PAs in our measurements at 3 bands appear to
be consistent with the PA measured by \citet{SZ91}. Those PAs are
roughly parallel to the radio axis which is at PA $\sim$ 91\degr\
\citep{Re99}.

\section{Conclusions}

We have presented the results of multi-band near-infrared polarimetry
of three quasars which have essentially unpolarized broad lines in the
optical.  The optical polarization is likely due to scattering
interior to the BLR, and near-IR polarization would also be from
the same mechanism, at least based on the similar polarization PA
observed.  In this case, the overall optical-near-IR polarized
flux is most simply interpreted as an electron-scattered version of
the radiation from interior to the BLR. This would provide the
measurement of the intrinsic spectral shape of the big blue bump
emission, revealing its shape in the near-IR for the first
time. 

The observed polarized flux slope in the optical to near-IR looks
quite blue compared to the shape in the optical alone. The near-IR
slope observed in Ton202 seems to be as blue as the long wavelength
limit of simple multi-temperature black-body disks and is consistent
with the bare disk atmosphere model spectrum which we have computed
for illustrating the comparison.  However if the disk is to be
truncated at the outer radius where the disk becomes nominally
self-gravitating, the model spectrum becomes much bluer in the
near-IR, deviating from the observed near-IR polarized flux data.  The
future data with higher S/N for many object would be able to
characterise the spectral shape for better comparisons with the
models.

With the optical polarized flux data, specifically the discovery of
the Balmer edge in absorption in several objects, and now this
preliminary demonstration of the near-IR polarized flux as an
indication of the blue underlying continuum, we seem to be sketching
in the true spectrum of the quasar big blue bump.  The results so far
look more intelligible physically than the total flux spectra.  While
the quasi-static Shakura-Sunyaev type disk is known to have many
problems, evidence seems to be accumulating that something at least
apparently resembling a classical disk, i.e. an optically thick,
thermally emitting disk, may exist in quasars.  We are exploring these
ideas with much more accurate polarimetry data.

We also have presented the $K$ band polarimetry of other quasars and
radio galaxies. The overall polarization properties seem to be
understandable in a picture where two polarization components compete,
namely the scattering in an equatorial region interior or close to the
BLR and the scattering in a polar region quite outside of the BLR.

\section*{Acknowledgements}

The United Kingdom Infrared Telescope (UKIRT) is operated by the Joint
Astronomy Centre on behalf of the U.K. Particle Physics and Astronomy
Research Council.  We are grateful to Chris Davis for kindly helping
us to obtain the data through the UKIRT service programme.  We thank
Zhaohui Shang and Beverley Wills for providing the electronic data of
their spectra for calibrating our data.  We thank the Department of
Physical Sciences, University of Hertfordshire for providing IRPOL2
for the UKIRT.  This research has made use of the NASA/IPAC
Extragalactic Database (NED) which is operated by the Jet Propulsion
Laboratory, California Institute of Technology, under contract with
the National Aeronautics and Space Administration. This research has
also made use of the SIMBAD database, operated at CDS, Strasbourg,
France. The work by OB was supported in part by NSF grant 
AST-0307657.


\label{lastpage}


\begin{thebibliography}{99}


\bibitem[\protect\citeauthoryear{Antonucci}{1988}]{An88}
Antonucci, R. 1988, in Supermassive Black Holes, ed. M. Kafatos
(Cambridge: Cambridge Univ. Press), 26

\bibitem[\protect\citeauthoryear{Antonucci \& 
Barvainis}{1990}]{AB90} Antonucci R., Barvainis R., 1990, 
ApJ, 363, L17 
 
\bibitem[\protect\citeauthoryear{Antonucci}{1999}]{An99}
Antonucci, R. 1999, in High Energy Processes in Accreting Black Holes, 
ASP Conf. Ser. 161, 193

\bibitem[\protect\citeauthoryear{Antonucci}{1993}]{An93} 
Antonucci R., 1993, ARA\&A, 31, 473

\bibitem[\protect\citeauthoryear{Berriman et al.}{1990}]{Be90}
Berriman, G., Schmidt, G. D., West, S. C., Stockman, H. S., 1990, 
ApJS, 74, 869

\bibitem[\protect\citeauthoryear{Bonning et 
al.}{2005}]{Bo05} Bonning E.~W., Shields G.~A., Salviander 
S., McLure R.~J., 2005, ApJ, 626, 89 
 
\bibitem[\protect\citeauthoryear{Brindle et 
al.}{1990}]{Br90} Brindle C., Hough J.~H., Bailey J.~A., 
Axon D.~J., Ward M.~J., Sparks W.~B., McLean I.~S., 1990, MNRAS, 244, 604 
 
\bibitem[\protect\citeauthoryear{Brown \& McLean}{1977}]{BM77}
Brown, J. C, McLean, I. S., 1977, A\&A, 57, 141

\bibitem[\protect\citeauthoryear{Cardelli, Clayton, \& 
Mathis}{1989}]{Ca89} Cardelli J.~A., Clayton G.~C., Mathis 
J.~S., 1989, ApJ, 345, 245 
 
\bibitem[\protect\citeauthoryear{Cohen et al.}{1999}]{Co99} 
Cohen M.~H., Ogle P.~M., Tran H.~D., Goodrich R.~W., Miller J.~S., 1999, 
AJ, 118, 1963 

\bibitem[\protect\citeauthoryear{di Serego Alighieri et 
al.}{1996}]{diS96} di Serego Alighieri S., Cimatti A., 
Fosbury R.~A.~E., Perez-Fournon I., 1996, MNRAS, 279, L57 

\bibitem[\protect\citeauthoryear{Fernini et 
al.}{1993}]{Fe93} Fernini I., Burns J.~O., Bridle A.~H., 
Perley R.~A., 1993, AJ, 105, 1690 
 
\bibitem[\protect\citeauthoryear{Goodrich \& Miller}{1988}]{GM88}
Goodrich, R. W., Miller, J. S. 1988, ApJ, 331, 332

\bibitem[\protect\citeauthoryear{Goodrich \& 
Miller}{1994}]{Go94} Goodrich R.~W., Miller J.~S., 1994, 
ApJ, 434, 82

\bibitem[\protect\citeauthoryear{Goodman}{2003}]{Go03} 
Goodman J., 2003, MNRAS, 339, 937 

\bibitem[\protect\citeauthoryear{Heiles}{2000}]{He00} Heiles 
C., 2000, AJ, 119, 923 
  
\bibitem[\protect\citeauthoryear{Hubeny et al.}{2000}]{Hu00} Hubeny,
I., Agol, E., Blaes, O.,  Krolik, J. H., 2000, ApJ, 533, 710

\bibitem[\protect\citeauthoryear{Hubeny et al.}{2001}]{Hu01} Hubeny,
I., Blaes, O.,  Krolik, J. H., Agol, E., 2001, ApJ, 559, 680

\bibitem[\protect\citeauthoryear{Hurt et al.}{1999}]{Hu99} 
Hurt T., Antonucci R., Cohen R., Kinney A., Krolik J., 1999, ApJ, 514, 579 
 
\bibitem[\protect\citeauthoryear{Jackson \& 
Rawlings}{1997}]{JR97} Jackson N., Rawlings S., 1997, MNRAS, 
286, 241 

\bibitem[\protect\citeauthoryear{Jannuzi \& 
Elston}{1991}]{JE91} Jannuzi B.~T., Elston R., 1991, ApJ, 
366, L69 

\bibitem[\protect\citeauthoryear{Kaspi et al.}{2005}]{Ka05} 
Kaspi S., Maoz D., Netzer H., Peterson B.~M., Vestergaard M., Jannuzi 
B.~T., 2005, ApJ, 629, 61 
 
\bibitem[\protect\citeauthoryear{Kishimoto et 
al.}{2001}]{Ki01} Kishimoto M., Antonucci R., Cimatti A., 
Hurt T., Dey A., van Breugel W., Spinrad H., 2001, ApJ, 547, 667 
 
\bibitem[\protect\citeauthoryear{Kishimoto, Antonucci, \& 
Blaes}{Kishimoto et al.}{2003}]{Ki03} Kishimoto M., Antonucci R., Blaes O., 
2003, MNRAS, 345, 253 
 
\bibitem[\protect\citeauthoryear{Kishimoto et 
al.}{2004}]{Ki04} Kishimoto M., Antonucci R., Boisson C., 
Blaes O., 2004, MNRAS, 354, 1065 
 
 \bibitem[\protect\citeauthoryear{Koratkar \& Blaes}{1999}]{KB99}
Koratkar, A., Blaes, O. 1999, PASP, 111, 1

\bibitem[\protect\citeauthoryear{Korhonen \& 
Reiz}{1986}]{KR86} Korhonen T., Reiz A., 1986, A\&AS, 64, 
487 

\bibitem[\protect\citeauthoryear{Laing}{1981}]{La81} Laing 
R.~A., 1981, MNRAS, 195, 261  

\bibitem[\protect\citeauthoryear{Laing et al.}{1994}]{La94} 
Laing R.~A., Jenkins C.~R., Wall J.~V., Unger S.~W., 1994, ASPC, 54, 201 

\bibitem[\protect\citeauthoryear{Malkan}{1983}]{Ma83} Malkan 
M.~A., 1983, ApJ, 268, 582 
 
\bibitem[\protect\citeauthoryear{Malkan}{1989}]{Ma89} Malkan 
M., 1989, in Theory of Accretion Disks, Proceedings of a
NATO Advanced Research Workshop, ed. F. Meyer (Dordrecht: Kluwer), 
19

 
\bibitem[\protect\citeauthoryear{Moore \& 
Stockman}{1984}]{MS84} Moore R.~L., Stockman H.~S., 1984, 
ApJ, 279, 465 
 
\bibitem[\protect\citeauthoryear{Miller, Robinson \& Goodrich}
{Miller et al.}{1988}]{MRG88} Miller, J. S., Robinson, L. B., Goodrich,
R. W., 1988, in Instrumentation for Ground-Based Astronomy,
ed. L. B. Robinson (New York, Springer), p. 157

\bibitem[\protect\citeauthoryear{Morganti, Killeen, \& 
Tadhunter}{1993}]{Mo93} Morganti R., Killeen N.~E.~B., 
Tadhunter C.~N., 1993, MNRAS, 263, 1023 
 
\bibitem[\protect\citeauthoryear{Reid, Kronberg, \& 
Perley}{1999}]{Re99} Reid R.~I., Kronberg P.~P., Perley 
R.~A., 1999, ApJS, 124, 285 
 
\bibitem[\protect\citeauthoryear{Rusk \& 
Seaquist}{1985}]{RS85} Rusk R., Seaquist E.~R., 1985, AJ, 
90, 30 
 
\bibitem[\protect\citeauthoryear{Shang et al.}{2003}]{Sh03} 
Shang Z., Wills B.~J., Robinson E.~L., Wills D., Laor A., Xie B., Yuan J., 
2003, ApJ, 586, 52 

\bibitem[\protect\citeauthoryear{Shields}{1978}]{Sh78} 
Shields G.~A., 1978, Natur, 272, 706 
  
\bibitem[\protect\citeauthoryear{Sitko \& Zhu}{1991}]{SZ91} 
Sitko M.~L., Zhu Y., 1991, ApJ, 369, 106 

\bibitem[\protect\citeauthoryear{Simmons \& 
Stewart}{1985}]{SS85} Simmons J.~F.~L., Stewart B.~G., 1985, 
A\&A, 142, 100  

\bibitem[\protect\citeauthoryear{Smith, Schmidt, \& 
Allen}{1993}]{Sm93} Smith P.~S., Schmidt G.~D., Allen R.~G., 
1993, ApJ, 409, 604 
 

\bibitem[\protect\citeauthoryear{Smith et al.}{2002}]{Sm02}
Smith, J. E., Young, S., Robinson, A., Corbett, E. A., Giannuzzo, M. E., 
Axon, D. J., Hough, J. H. 2002, MNRAS, 335, 773


\bibitem[\protect\citeauthoryear{Smith et al.}{2005}]{Sm05} 
Smith J.~E., Robinson A., Young S., Axon D.~J., Corbett E.~A., 2005, MNRAS, 
359, 846 
 
\bibitem[\protect\citeauthoryear{Smith et al.}{2004}]{Sm04} 
Smith J.~E., Robinson A., Alexander D.~M., Young S., Axon D.~J., Corbett 
E.~A., 2004, MNRAS, 350, 140 

\bibitem[\protect\citeauthoryear{Stockman, Angel, \& 
Miley}{1979}]{St79} Stockman H.~S., Angel J.~R.~P., Miley 
G.~K., 1979, ApJ, 227, L55 
 
\bibitem[\protect\citeauthoryear{Stockman, Moore \& Angel}{Stockman et
al.}{1984}]{St84} Stockman, H. S., Moore, R. L., Angel, J. R. P. 1984,
ApJ, 279, 485

\bibitem[\protect\citeauthoryear{Stockton \& 
MacKenty}{1987}]{SM87} Stockton A., MacKenty J.~W., 1987, 
ApJ, 316, 584 

\bibitem[\protect\citeauthoryear{Tokunaga \& 
Vacca}{2005}]{TV05} Tokunaga A.~T., Vacca W.~D., 2005, PASP, 
117, 421

\bibitem[\protect\citeauthoryear{Ulvestad \& 
Antonucci}{1988}]{UA88} Ulvestad J.~S., Antonucci R.~R.~J., 
1988, ApJ, 328, 569 
 
\bibitem[\protect\citeauthoryear{Whittet et 
al.}{1992}]{Wh92} Whittet D.~C.~B., Martin P.~G., Hough 
J.~H., Rouse M.~F., Bailey J.~A., Axon D.~J., 1992, ApJ, 386, 562 
 
\bibitem[\protect\citeauthoryear{Wills et al.}{1992}]{Wi92}
Wills, B. J., Wills, D., Breger, M., Antonucci, R. R. J., Barvainis,
R. 1992, ApJ, 398, 454

\bibitem[\protect\citeauthoryear{Young et al.}{1998}]{Yo98} 
Young S., Hough J.~H., Axon D.~J., Fabian A.~C., Ward M.~J., 1998, MNRAS, 
294, 478

\bibitem[\protect\citeauthoryear{Young et al.}{1999}]{Yo99}
Young, S., Corbett, E. A., Giannuzzo, M. E., Hough, J. H., Robinson, A., 
Bailey, J. A., Axon, D. J. 1999, MNRAS, 303, 227



\end{thebibliography}
\end{document}